# Optical Pulling Force in Carbon Nanotubes: Manifestation of Nonlocal Conductivity

Tomer Berghaus [1,*], Touvia Miloh [1] and Gregory Ya. Slepyan [2,**]

[1] *School of Mechanical Engineering, Tel Aviv University, Tel Aviv 69978, Israel;*

[2] *School of Electrical Engineering, Tel Aviv University, Tel Aviv 69978, Israel;*

We develop a new theory of an optical force exerted on a carbon nanotube (CNT) with a nonlocal conductivity. The optical force is expressed in terms of the surface current density and the axial electric field on the CNT surface. To determine these quantities, we employ an integral-equation-based approach in terms of the current density. The analysis is constructed for a finite-length cylindrical CNT by rigorously accounting for edge effects. In addition to numerical solutions of the integral equation, we obtain an approximate analytical expression for the optical force acting on the CNT, which shows good agreement with numerical simulations. We also demonstrate the existence of some frequency ranges in which the optical force becomes negative, corresponding to the optical pulling effect. Such a pulling behavior is shown to originate from the nonlocality of the conductivity and to vanish in the local limit. This work advances theoretical understanding of optomechanical interactions in finite-length low-dimensional conductors and clarifies the role of spatial dispersion in the emergence of optical pulling forces.

*Introduction* – Studying forces and torques arising from the interaction of light with matter is essential for understanding fundamental aspects of electrodynamics and has opened the way to a broad range of non-destructive and non-contact manipulation techniques for micro- and nanoscale particles in physics [1–12], biology [6], chemistry [2,13], and medicine [14]. One of the earliest milestones in this field were the pioneering observations of electromagnetic radiation pressure by Lebedev [15] and Nichols with Hullin [16] (both in 1901). The modern development of this research area can naturally be divided into two principal stages. The first stage began with Ashkin's pioneering works [17,18], which focused on micrometer-sized dielectric or metallic particles manipulated by strongly focused laser beams in the visible frequency range. During this period, theoretical models of optical forces were developed within the realm of classical electrodynamics. Material response was described phenomenologically by means of using constitutive parameters such as permittivity and conductivity. Typically, the particle size was much smaller than the wavelength, and scattering was treated in the Rayleigh regime by employing the dipole approximation.

Within this framework, two fundamental mechanisms of optical forces were identified. The first is the gradient force, governed by the spatial gradient of the field intensity, which is present even in lossless particles [17]. The second is the scattering force, associated with momentum transfer from the electromagnetic field, and exists in absorbing media [17]. In most conventional configurations, optical forces act in the direction of wave propagation (optical pushing). One of the practical outcomes of this stage was the establishment of optical tweezers [13, 19-23] and modern force microscopy techniques [24].

Recent advances in nanotechnology, have initiated a second stage in the physics of optical forces. First, the accessible frequency range was expanded towards the infrared and terahertz domains. Second, the range of physical mechanisms governing optical forces has been significantly broadened. At the nanoscale, the origins of optical forces are no longer reducible solely due to Lorentz polarization of atomic ensembles or Ohmic conduction of free electrons in a crystal lattice. In many cases, it is practically insufficient to employ a purely local classical description, expressed in terms of effective permittivity and permeability.

The conventional picture has therefore been extended to include effects of chirality [25,26], plasmonic resonances in noble metals [27,28], optical gain and gain–loss balance within a single particle [29,30], spatial dispersion (nonlocality of current or dipole response at a given point) [31], composite configurations such as Janus particles [32,33], excitonic effects in semiconductor nanostructures [34,35], optical nonlinearities [36], metamaterials [37, 38] and spinning particles [39]. A major conceptual advance of this stage was the prediction and realization of counterintuitive negative optical forces acting in the opposite direction of wave propagation (optical pulling) [40–51]. A negative optical torque was also predicted [52]. From a practical perspective, these new developments have extended the use of optical manipulation techniques into lower-frequency regimes as well as to a wide area of novel nanoscale systems.

Despite the substantial progress, several fundamental issues remain unresolved. Optical forces are commonly calculated by integrating the electromagnetic Maxwell stress tensor over a closed surface enclosing the object [53]. There exist several formulations for the macroscopic electromagnetic stress tensor [54–58]. While these formulations yield identical results when the integration surface lies entirely in a homogeneous, unbounded medium, they can lead to different predictions when the surface intersects material boundaries, such as those of condensed matter [59]. The selection of the appropriate tensor formulation in such cases remains an open question.

Another common assumption is using a point-like scatterer model, in which the incident field is considered practically unperturbed by the presence of the particle (Rayleigh's approximation). However, the validity of this approximation is not always guaranteed, even when the particle size is small compared to the wavelength [51]. Attempts to refine the theory within the Rayleigh regime, by accounting, for example, for the longitudinal field component in tightly focused beams, have been reported in [60].

In recent years, a growing attention has been devoted to optical forces in carbon-based nanostructures, such as graphene and carbon nanotubes (CNTs), owing to their exceptional electronic and optical properties [61–68]. In this work, we present an original study of optical forces acting on a CNT beyond the conventional dipole approximation, by taking into account the effect of the nonlocal surface conductivity. We demonstrate that a realistic description of the problem requires considering a finite-length CNT model, which introduces the fundamental challenge of accurately capturing edge (end) effects. Within this framework, we also demonstrate the emergence of an optical pulling nature in CNTs, which originates from the nonlocal nature of the conductivity and which has no analog in the traditional treatment of locally responding structures.

*Non-local CNT conductivity* – The model of nonlocal conductivity of a CNT was developed in [69], covering the wide frequency range (from THz to visible light inclusively). We consider a finite-length conductive zigzag CNT ($m$,0) with $m$=3$s$, where $s$ is an integer value and simply employ the Kubo formalism for low-energy excitations, constrained by the condition $E < 1\,\text{eV}$. For this purpose, we adapt it for the case of a pseudo-spin electron liquid, in a similar manner to the approach used for a monolayer graphene in [70, 71]. To this end, we consider the electronic states near the K-points, assuming a linear approximation for $\varepsilon_{\mathbf{p}} = \pm v_F \left|\mathbf{p} - \mathbf{p}_F\right|$ ($v_F$ is the Fermi velocity). In addition, we consider a collision-free electron motion for low temperatures ($T << \mu$, $\mu$ is the electrochemical potential). In contrast to graphene [70, 71], for a CNT we also consider the transverse electron confinement (i.e., the azimuthal component of the electron momentum is discrete while the longitudinal component is continuous).

The electromagnetic fields and the induced currents in the case of spatial dispersion can be expressed as a superposition of traveling waves $\exp\left[i\left(qz - \omega t\right)\right]$ with wavenumber $q$, which are propagating in opposite directions. Far from the CNT terminations (ends), their interaction can be accurately modeled by using the corresponding conventional infinite-length CNT approximation. Reflections from the extreme ends of the CNT and accounting for nonlocal effects, are incorporated in the present formulation, by applying the so-called "additional" boundary conditions [72] (for detailed calculations see Supplemental Material [73]).

The conductivity in the momentum space can be defined as $\sigma_{zz}\left(\omega,q\right) = i\left[\Pi\left(\omega,q\right) - \Pi\left(0,q\right)\right]/\omega$ , where

$$\Pi(\omega,q)=\sum_{mn}\sum_{p}\frac{F(\varepsilon_{m,p+q})-F(\varepsilon_{np})}{\left[\varepsilon_{np}-\varepsilon_{m,p+q}-\hbar(\omega+i0)\right]}\left|\langle\hat{j}_z\rangle_{mn,p,p+q}\right|^2 \quad (1)$$

denotes the polarizability and $\langle\hat{j}_z\rangle_{mn,p,p+q}$ represent the matrix elements of the operators of the current density. The symbol $\omega+i0$ is conventionally associated with collisionless attenuation. The integers sum over $p$ is a superposition over the states with different momenta and is a conventional form of indicating an integration over the BZ zone. It is also convenient to split the total nonlocal conductivity into inter and intra components, namely, $\sigma_{zz}(\omega,q)=\sigma_{zz}^{\text{inter}}(\omega,q)+\sigma_{zz}^{\text{intra}}(\omega,q)$. The first term with $m\neq n$ in (1), corresponds to interband transitions while the second to intraband transitions (also known as the Drude component in conductivity). In addition, we note that the spatial dispersion can be also approximated by Taylor series with two first order terms (the second term vanishes due to the symmetry of CNT). We have

$$i\omega^{-1}\Pi(\omega,q)\approx i\omega^{-1}\left[\Pi(\omega,0)+\frac{1}{2}\frac{\partial^2\Pi(\omega,q)}{\partial q^2}\bigg|_{q=0}\cdot q^2\right] \quad (2)$$

Returning next to the position space, we perform an inverse transformation $q\to\partial/\partial z$, which suggests that the total current can be expressed as the sum of local and non-local components, i.e., $j_z=\sigma_{zz}(\omega,0)E_z+j_{z,\text{Nonloc}}$, where

$$j_{z,\text{Nonloc}}=-\xi(\omega)\frac{\partial^2E_z}{\partial z^2} \quad (3)$$

with the factor of nonlocality in (3) given by $\xi(\omega)=\frac{1}{2}\frac{\partial^2\sigma_{zz}(\omega,q)}{\partial q^2}\Big|_{q=0}$.

Note that the cylindrical CNT has a finite radius, which corresponds to rather small values of $m$ ($m<60$). The main contribution to the conductivity is provided by the two lines crossing corresponding to the Fermi points, related to the different valleys of the BZ zone. The next step is to transform the integration over $p_z$ to that over energy, and then carry it out over the range $0<\varepsilon<\infty$, resulting in

$$\sigma_{zz}^{\text{inter}}(\omega)=\frac{ie^2\hbar\omega v_F}{\pi^2R_{CN}}\int_0^\infty\frac{F(-\varepsilon)-F(\varepsilon)}{\left[\hbar^2(\omega+i0)^2-4\varepsilon^2\right]^2}\frac{d\varepsilon}{\varepsilon} \quad (4)$$

The integral in Eq. (4) is like the graphene conductivity integral derived in Ref. [70], but it is not identical. The difference between the two expressions, is due to the additional energy-dependent factor in the denominator of the integrand. This parameter originates from the distinct dimensionality of the electronic states: in a CNT, the momentum summation reduces to an integration along two discrete one-dimensional subbands associated with the Fermi points, whereas in graphene it extends over the continuous two-dimensional Brillouin zone. In a similar manner, the complementary intraband conductivity can be written as

$$\sigma_{zz}^{\text{intra}}(\omega)=-\frac{e^2v_F}{i\pi^2\hbar\omega R_{CN}}\int_{-\infty}^{\infty}\frac{\varepsilon}{|\varepsilon|}\frac{\partial F}{\partial\varepsilon}d\varepsilon \quad (5)$$

The inter-band conductivity component dominates at high frequencies, while the intra-band component is significant at low frequencies.

The non-locality factor in (3), can be expressed as

$$\xi(\omega) = \frac{1}{2} v_F^2 \frac{\partial^2 \sigma_{zz}(\omega)}{\partial \omega^2} \ . \tag{6}$$

(for detailed calculations, see Supplemental Material [73]). The non-locality factor is a complex value, which contributes to the collisionless relaxation of a CNT too. It satisfies the Kramers-Kroning relation and can also be splitted into intraband and interband components, in a similar manner to conductivity. Our model remains accurate for $R_{CN}$ <30nm, 10nm<$L$<1mm, $f$< 300 THz, $\mu$<0.5 eV [69].

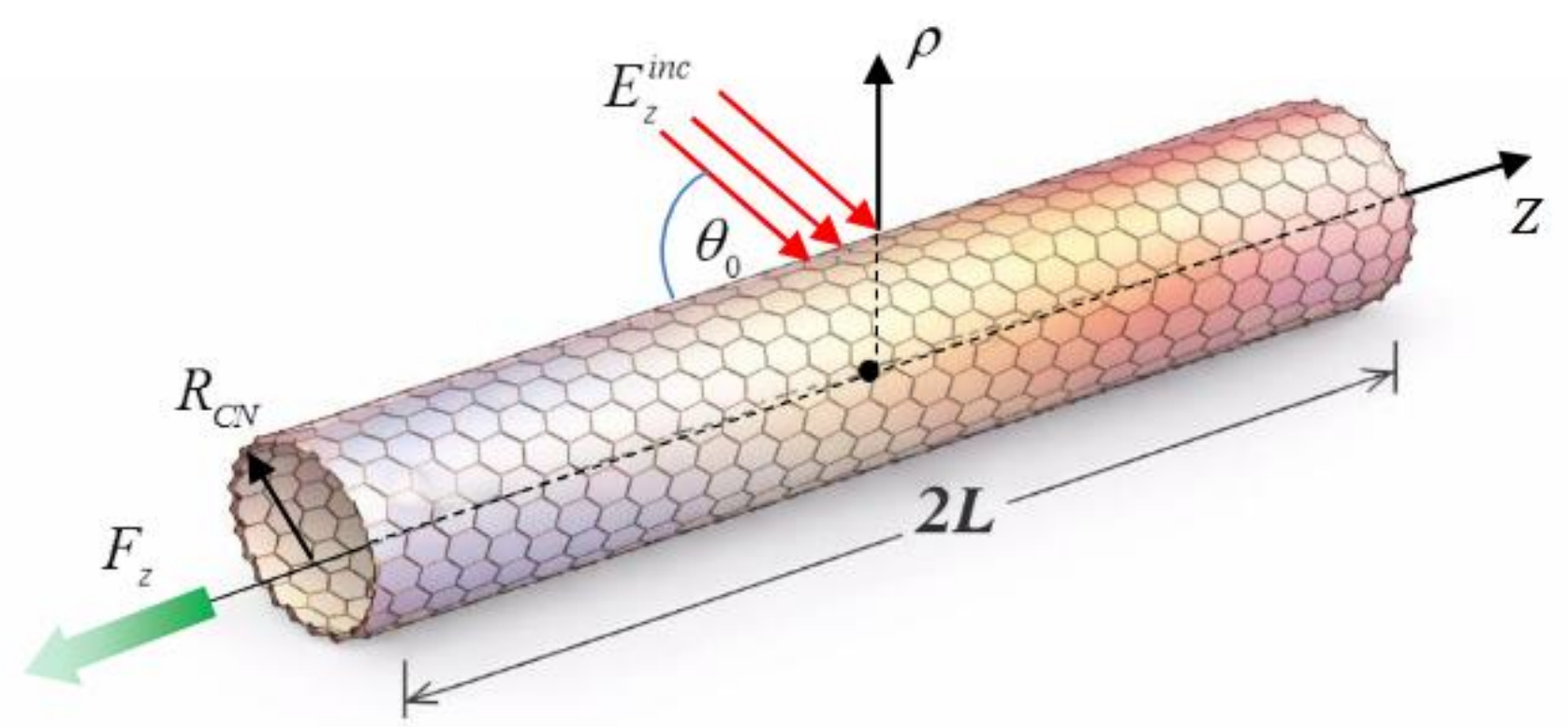

FIG 1. Configuration of a zigzag CNT with unclosed ends. CNT is excited by plane wave with angle of incidence $\theta_0$ counted from the longitudinal axis. The origin of cylindrical coordinate system is placed at the central point of CNT axis. The dominative component of the optical force is axial one (it is shown the direction of pulling case).

*Optical force* - Let us consider the oblique incidence of a plane wave $E_z^{inc}(z) = E_0 \sin\theta_0 e^{ikz\cos\theta_0}$, on a freely suspended CNT, where $\theta_0$ denotes the angle of incidence. All forms of stress tensors are principally equal. For CNTs, the circle-averaged optomechanical force is evaluated by integrating the circle-averaged Maxwell stress tensor (MST) over the nanotube surface, i.e., $\mathbf{F} = \int_S \vec{\mathbf{T}} \cdot \mathbf{n} dS = 2\pi R_{CN} \int_{-L}^{L} \left( \vec{\mathbf{T}}^{(+)} - \vec{\mathbf{T}}^{(-)} \right) \mathbf{e}_r dz$. Here $\vec{\mathbf{T}}^{(\pm)}$ denote the values of the stress tensor on the external and internal sides of the CNT, while $\mathbf{e}_r$ is the radial unit vector. For MST calculation, we use EM-field averaged with respect to the infinitely small element of CNT surface. As a result, any physical field will be exchanged by its envelope with respect to the oscillations with respect to the scale of the crystal lattice (macroscopic field). The macroscopic field satisfies Maxwell equations with effective boundary conditions [72], which couple the electric and magnetic field components, both on the internal and external sides of CNT (for details see Supplemental Material [73]).

The circle-averaged MST can also be expressed as $\vec{\mathbf{T}} = 0.5 \cdot \mathrm{Re}\left[ \left( \varepsilon_0 \mathbf{E}\mathbf{E}^* + \mu_0 \mathbf{H}\mathbf{H}^* \right) - 0.5 \cdot \left( \varepsilon_0 |\mathbf{E}|^2 + \mu_0 |\mathbf{H}|^2 \right) \mathbf{I} \right]$, with $\mathbf{I}$ denoting the unit tensor. Since the CNT is slender, the field is azimuthally symmetric and has a form $\{\mathbf{E};\mathbf{H}\} = \{E_r, E_z; H_\varphi\}$, and thus the MST renders

$$\ddot{\mathbf{T}}^{(\pm)} = \frac{1}{4}\mathrm{Re}\begin{pmatrix} \varepsilon_0\left|E_r^{(\pm)}\right|^2 - \varepsilon_0\left|E_z^{(\pm)}\right|^2 - \mu_0\left|H_\varphi^{(\pm)}\right|^2 & 0 & 2\varepsilon_0 E_r^{(\pm)}\left(E_z^{(\pm)}\right)^* \\ 0 & \mu_0\left|H_\varphi^{(\pm)}\right|^2 - \varepsilon_0\left|E_r^{(\pm)}\right|^2 - \varepsilon_0\left|E_z^{(\pm)}\right|^2 & 0 \\ 2\varepsilon_0 E_z^{(\pm)}\left(E_r^{(\pm)}\right)^* & 0 & \varepsilon_0\left|E_z^{(\pm)}\right|^2 - \varepsilon_0\left|E_r^{(\pm)}\right|^2 - \mu_0\left|H_\varphi^{(\pm)}\right|^2 \end{pmatrix} \quad (7)$$

As one can see, the optical force in CNT axially directed and defined by the components $T_{\rho z}$, $T_{z\rho}$ (all other MST components responsible for CNT deformations only). It is worth noting that using the local model may lead to unphysical results, since the integration of the radial component of the electric field in Eq. (7) diverges due to nonintegrable singularities in the diagonal elements. In contrast, when nonlocal effects are taken into account, these singularities are removed, and the integration in Eq. (7) becomes convergent. Therefore, the inclusion of the spatial dispersion effects in the present formulation provides a physically consistent and mathematically well-posed description of optomechanical interactions in CNTs. Using the continuity of $E_z$ at the CNT surface, we obtain the following expression for the longitudinal optical force $F_z = 2\pi\varepsilon_0 R_{CN}\cdot\mathrm{Re}\int_{-L}^{L}\left(E_r^{(+)} - E_r^{(-)}\right)E_z^* dz$, where the discontinuity in the radial electric field is proportional to the surface charge density. By applying the continuity relation, and conducting integration by parts, we obtain

$$F_z = i\frac{\pi R_{CN}}{\omega}\int_{-L}^{L}\left(\frac{\partial E_z^*}{\partial z}j_z - \frac{\partial E_z}{\partial z}j_z^*\right)dz. \quad (8)$$

Equation (8) is the final result for the optical force, which is used in the numerical simulation of this paper.

*Integral equation* – Evaluation of the force in Eq. (8) requires determining the current density and electric field on the CNT surface. The integral equation approach has proven effective for solving such problems in CNT electrodynamics [74-81]. To this end, we employ the integral equation formulation for nonlocal conductivity developed in [69]. The electric field in (8) may be expressed through the current density and the integral equation may be solved numerically, while also allows obtaining a simple analytical approximate solution (both approaches will be considered in this Letter). The integral equation has the form of a Fredholm integral equation of the second kind and reads

$$j_z(z) - \frac{i\sigma_{zz}(\omega)\tilde{\alpha}^2(\omega)}{\omega\varepsilon_0}\int_{-L}^{L} j_z(s)K(z,s)ds = B(z) \quad (9)$$

with

$$B(z) = \sigma_{zz}(\omega)\tilde{\alpha}^2(\omega)\int_{-L}^{L} g(z,z')E_z^{inc}(z')dz' \quad (10)$$

where $\tilde{\alpha}(\omega) = \sqrt{\sigma_{zz}(\omega)/\xi(\omega)}$ is the special wavenumber defined by nonlocality (in the local limit $\tilde{\alpha}(\omega)\to\infty$, $g(z,z')\to\tilde{\alpha}^{-2}(\omega)\delta(z-z')$, $\delta(z-z')$ is Dirac delta-function). The kernel $g(z,z')$ in (10), denotes the 1D Green function of the 1D Helmholtz equation satisfying the Dirichlet boundary conditions $g(\pm L,z') = 0$. Accordingly, the kernel $K(z,s)$ in (9), is defined as

$$K(z,s) = \left(\frac{\partial^2}{\partial s^2} + k^2\right)\int_{-L}^{L} g(z,z')G(s-z')dz' \quad (11)$$

where $G(s-z') = G(R_{CN}, s-z')$ and

$$G(r,s-z)=R_{CN}\int_0^{2\pi}\frac{e^{ik\sqrt{r^2+R_{CN}^2-2rR_{CN}\sin(\phi)+(s-z)^2}}}{\sqrt{r^2+R_{CN}^2-2rR_{CN}\sin(\phi)+(s-z)^2}}d\phi \tag{12}$$

Equation (12) corresponds to the 3D Green function of the 3D Helmholtz equation in a free space, which is azimuthally integrated over the CNT radius. The two kernels $g(z,z')$ and $K(z,s)$ may be also written in different equivalent forms, which are most convenient for different purposes (see Supplemental Material [73]).

*Optical pulling force* – Using both numerical solution of the integral equation and the approximate analytical solution Eq. (13), we can determine the optical pulling force acting on the CNT. For the numerical solution of the integral equation (9), we approximate the integral in (9) by a finite sum with $N$ terms and select the $N$ testing points at the center of each segment. As a result, we obtain a system of linear equations of $N$-th order, which was solved via a MATLAB code of matrix inversion. However, one can see that the kernel $G(z'-s)$ is expressed in terms of a singular integral. Therefore, the change in the order of integration and differentiation is not appropriate (namely, the integral obtained in such a way is non - converging) and the resulting integral equation is ill-posed. Thus, its direct numerical solution using certain nonstandard methods, often results in some unphysical oscillations of the current density, especially near the

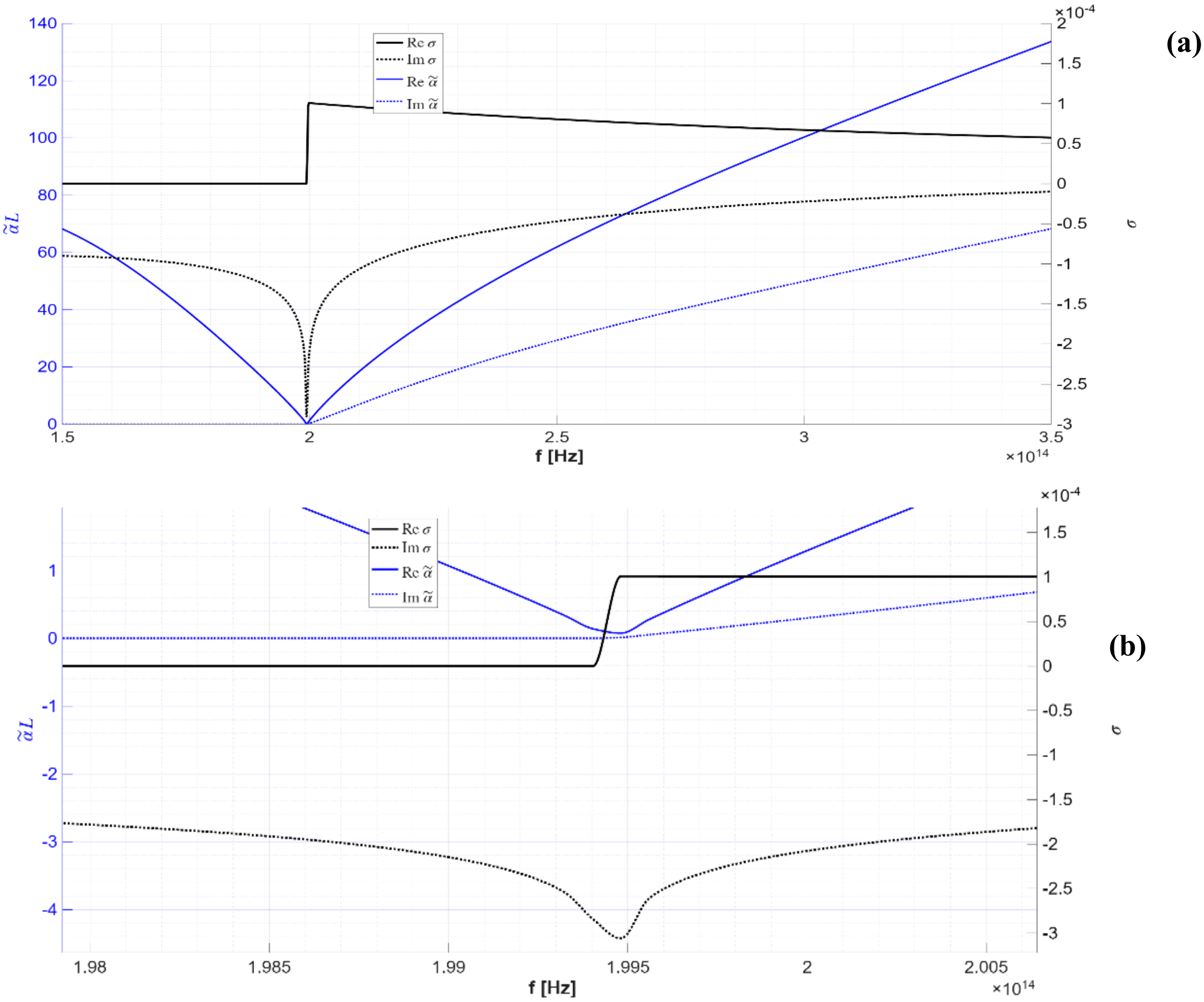

FIG.2. (a) Left axis: Nonlocal parameter (normalized by half of the length L=100nm). Right axis: axial conductivity $\sigma(\omega)$ as a function of frequency for a CNT (12,0); (b) Enlarged view at the vicinity of the resonance.

CNT termination points. Similar oscillations are well-known to exist also in the theory of classical microwave dipole antennas [82]. A renormalization approach aimed at mitigating such numerical artifacts, was proposed by Hanson et al. [82]. Although their technique is not directly applicable to our case, due to the differences in the equation's structure, the underlying principle remains CNT termination points. Similar oscillations are well-known to exist also in the theory of classical microwave dipole antennas [82]. A renormalization approach aimed at mitigating such numerical artifacts, was proposed by Hanson et al. [82]. Although their technique is not directly applicable to our case, due to the differences in the equation's structure, the underlying principle remains valuable. In order to adopt such an approach for a CNT, one can use the renormalized kernel defined in the Supplemental Material [73]. Yet another regularization technique, which is based on analytical integration of the singular component, is by using the additional theorem for Hankel functions (see details in Supplemental Material [73]).

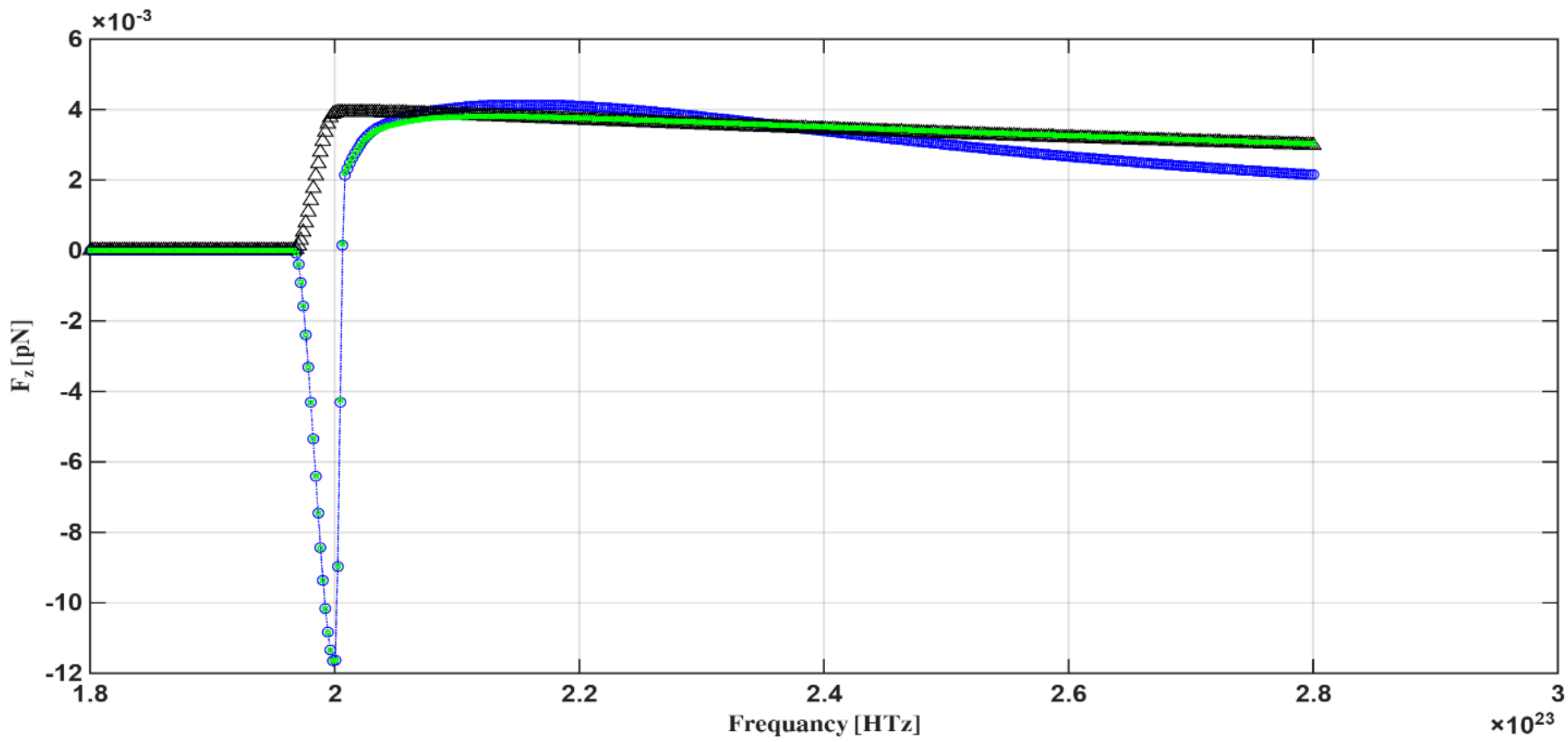

FIG. 3. Dependence of the optical force on frequency for a CNT (12,0) corresponding to the parameters: $E_0 = 10^7\,\mathrm{V/m}, R_{CN} = 0.47\mathrm{nm}, \mu{=}0.413\mathrm{eV}, L{=}100\mathrm{nm}, \theta_0{=}30^\circ$. Green squares - analytic relation; Black triangles - local conductivity model; Blue circles - non-local conductivity model. Number of segments for numerically solving the integral equation (9) and computing the current density, is $N$=411.

Recent technological advancements have enabled fabrication of CNTs with a wide range of lengths $2L$ [64, 83], from 10nm (ultra-short tubes) to 0.7–10 mm (exceptionally long tubes). We will consider the optical force for tubes with a reachable length within this range. Fig 3 shows the dependence of the optical force on frequency for the oblique incidence case of plane wave, demonstrating a good agreement between numerical simulations and the analytical solution given by Eq. (13). One can see the existence of a narrow peak of the negative force, namely a pulling optical force. The frequency of the peak (maximum) exactly corresponds to the condition

$$\frac{\hbar\omega}{2\mu} = 1 \tag{13}$$

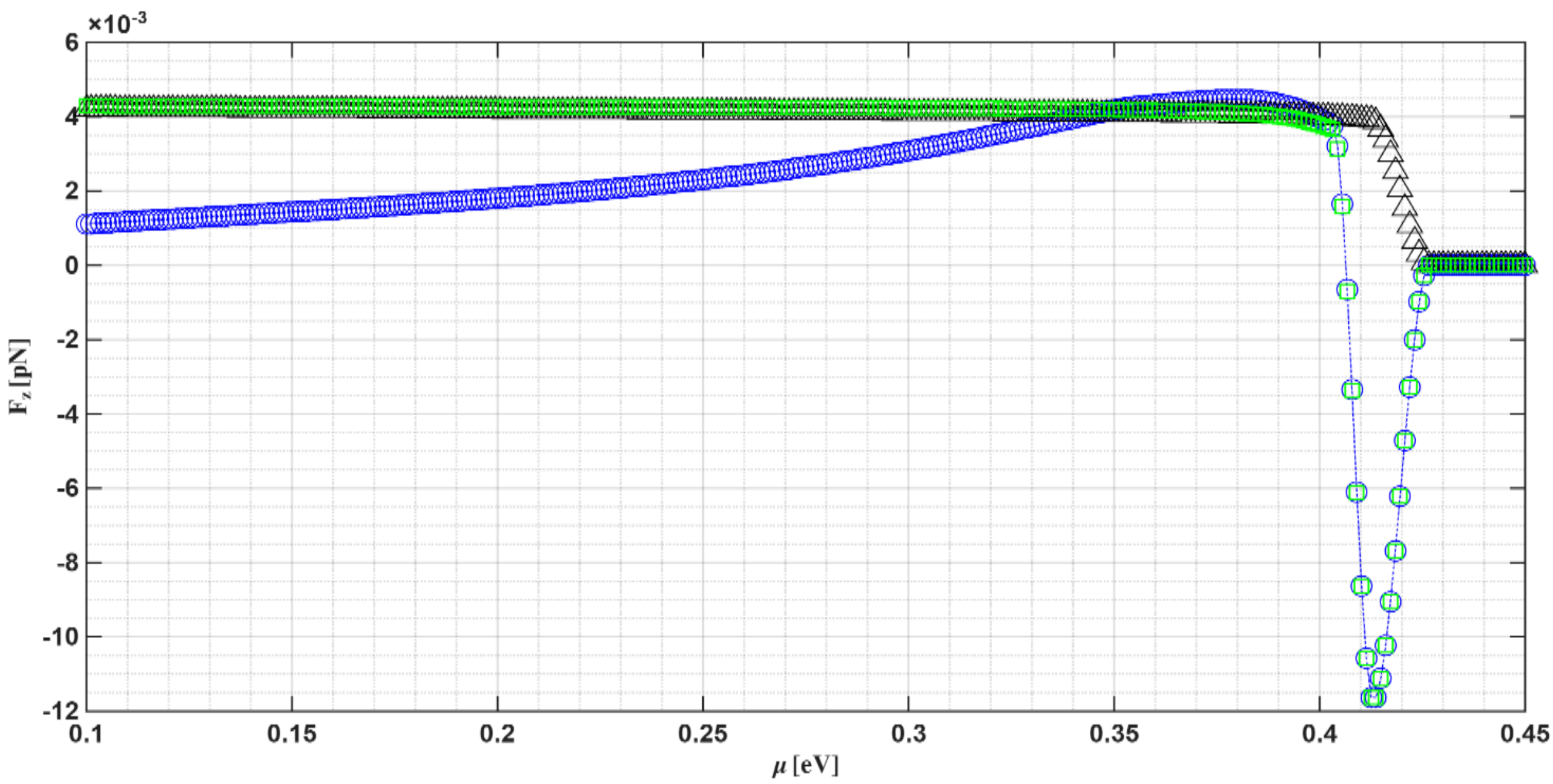

FIG.4. Dependence of the optical force on the electrochemical potential for a CNT (12,0): $E_0 = 10^7\,\mathrm{V/m}, R_{CN} = 0.47\mathrm{nm}, L=100\mathrm{nm}, \theta_0=30^\circ$, $f$=200THz. Green squares - analytic relation; Black triangles - local conductivity model; Blue Circles - non-local conductivity model. The number of segments for numerically solving the integral equation (9) and computing the current density, is $N$=161.

Equation (13) corresponds to the threshold frequency for the onset of collisionless attenuation, which manifests as a step-like feature in the conductivity (see Fig. 2). Figure 4 shows the dependence of the optical force on the electrochemical potential at a fixed frequency. A pronounced narrow negative peak (depending on μ) is observed, which is consistent with Eq. (13).

In Fig. 5 the dependence of force on the CNT length ($L$ is the half of the length) of a CNT for some given values of frequency and electrochemical potential. Note that we consider two principally different frequency ranges in numerical and analytical solutions. The first range corresponds to the first regime (intraband transitions) $\hbar\omega < 2\mu$ (Fig.5 (a)), while the second one corresponds to the second regime $\hbar\omega > 2\mu$ (Fig. 5 (b)). Both figures are constructed for the local limit of $\tilde{\alpha}(\omega) \to \infty$ too. It is evident that the force displays a qualitatively different behavior in the two regimes, were, for example, in regime 1 the force demonstrates oscillatory behavior. For a rather short CNT the force is negative, which corresponds to the optical pulling. The numerical simulations and the analytical solution (15) are in good agreement, also in determining the local limit. The computed optical force exhibits monotonic behavior and a linear dependence with respect to length. It means that the force density is uniformly distributed over the axis (the total force is given by the force density multiplied by the CNT length). Thus, we find that the effect of a pulling force is stipulated by non-locality arising solely from end effects. However, as shown, this situation is dramatically changed in the second regime. The force is positive even if non-local effects are taken into account, resulting in the common pushing (positive) force behavior. The curve obtained for the local case exactly agrees with the one which accounts for nonlocality, implying that edge effects can be ignored in this case. Note finally, that the selected cases considered in both regimes correspond to physically implementable values of CNT lengths. The values of force obtained in our simulations are realistic for observation and well-agreed with experimental data known for pushing regime [84, 85] (30-50 fN for incident power 50-100mW).

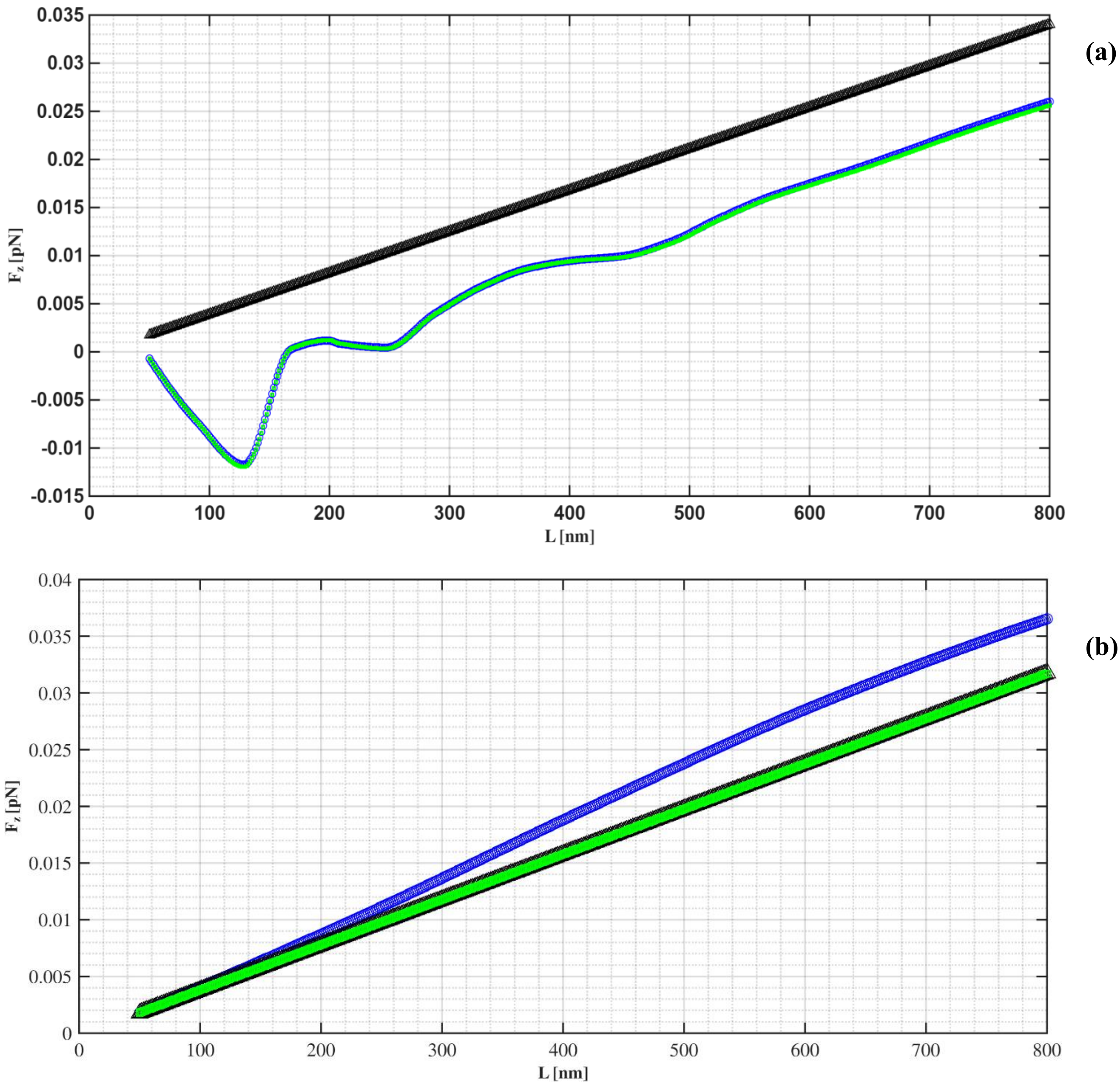

FIG. 5. Dependence of the optical force on the length of a CNT (12,0) for two frequencies: (a) $f$=200 THz and chemical potential 0.413eV and (b) $f$=215 THz: $E_0 = 10^7$ V/m, $R_{CN} = 0.47$nm, $\theta_0$=30°. Green squares - analytic relation; black triangles - local conductivity model; blue circles - non-local conductivity model. The number of segments for numerically solving the integral equation (9) and computing the current density, is $N$=411.

Dependence of the optical force on the angle of incidence for different values of CNT length is shown on Fig 6. CNTs exhibit the strong directional behavior of optical force – rather high dependence of the force on the angle of incidence (similar to the metal nanorodes [38]). It is important that pooling or pushing of the force takes place subject to the CNT length and is kept at the whole area of the angle variance.

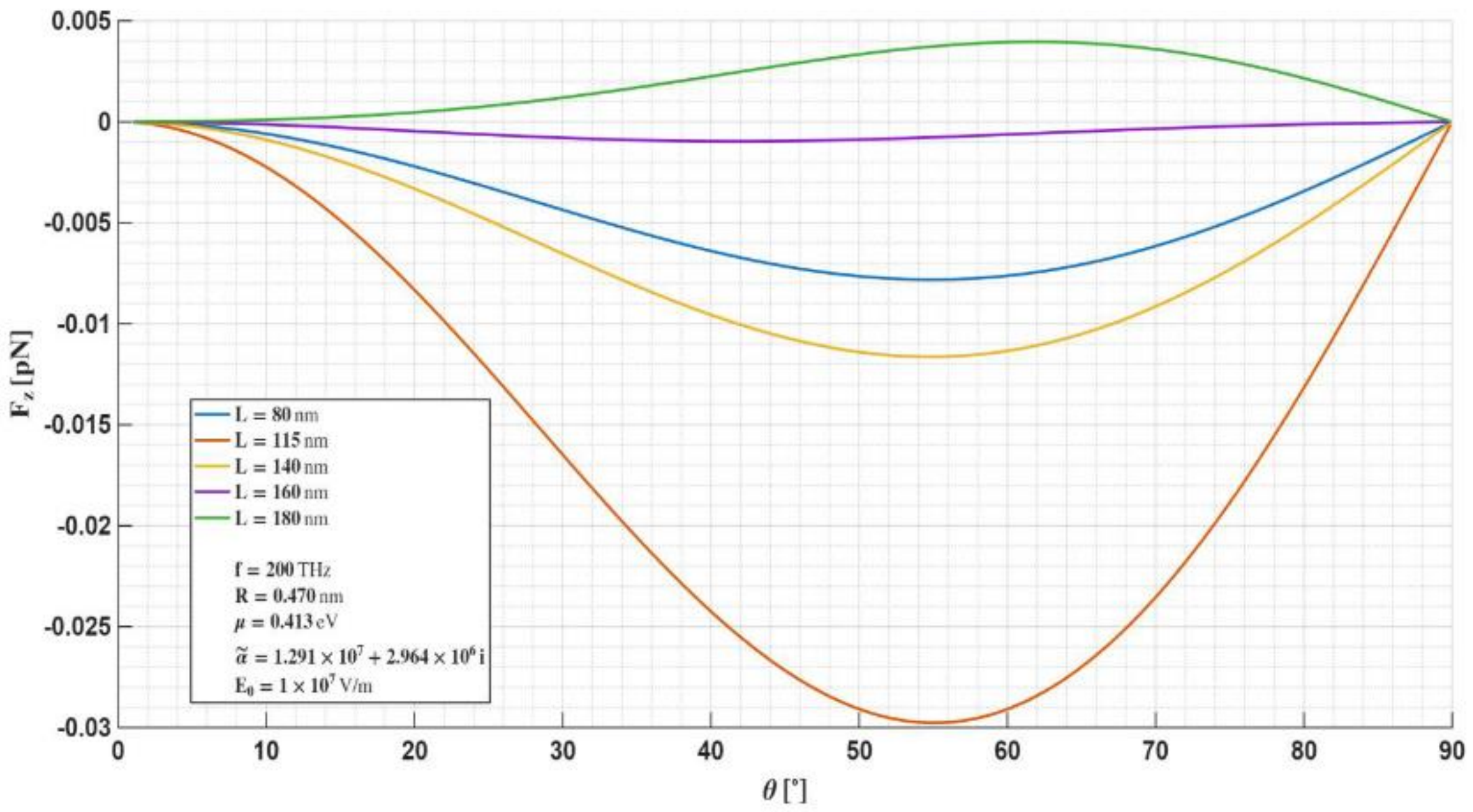

Fig. 6. Dependence of the optical force on the angle of incidence for different CNT lengths.

*Approximate analytical solution* – When accounting for non-locality, one can use an approximate solution by replacing

$$j_z(z) = j_z^{(0)}(z) = -B(z) \tag{14}$$

in Eq. (8). It corresponds to relation $E_z = E_z^{inc}$ at the surface of CNT. After elementary integration and by virtue of Eq. (8), one obtains the following result

$$F_z = A\,\mathrm{Re}\left\{\frac{\tilde{\alpha}^2(\omega)\sigma_{zz}(\omega)}{\tilde{\alpha}^2(\omega)-k^2\cos^2\theta_0}\left[L+\frac{\tilde{\alpha}(\omega)}{\sin\left(2\tilde{\alpha}(\omega)L\right)}\frac{\left(\cos\left(2\tilde{\alpha}(\omega)L\right)-\cos\left(kL\cos\theta_0\right)\right)}{\left(\tilde{\alpha}^2(\omega)-k^2\cos^2\theta_0\right)}\right]\right\} \tag{15}$$

where $A = 4\pi E_0^2 R_{CN} c^{-1}\sin^2\theta_0\cos\theta_0$. Eq. (15) demonstrates that non-locality effects qualitatively change the behavior of the optomechanical force. In the local limit $\tilde{\alpha}(\omega)\to\infty$ with fixed length the second term in Eq. (15) vanishes, while the first one gives

$$F_z = 4\pi L R_{CN} E_0^2 c^{-1}\sin^2\theta_0\cos\theta_0\,\mathrm{Re}\left(\sigma_{zz}(\omega)\right). \tag{16}$$

This limit corresponds to the homogeneous force density. The force becomes proportional to the CNT length, in agreement with Ashkin's-like model of optical tweezing. The second term in Eq. (15) is a non-monotonic and an alternating (oscillatory) quantity with respect to the length of the CNT, a dependence which is also manifested in optical pulling. The dispersion behavior of the optical force, various in the frequency range of existence or in the absence of collisionless attenuation ($\hbar\omega > 2\mu$ in the first case, and $\hbar\omega < 2\mu$ in the second one). It shows that the force in CNT is 'non-conservative' due to non-locality (it means that part of the provided energy is dissipated to produce mechanical action [12]).

*Conclusions and outlook* – In this work, we considered the simplest model of optical force in CNT: single achiral CNT in vacuum (air) excited via the plane wave. However, the developed approach may be rather simply generalized to the more complicated structures, such as (i) CNT suspended in gas or fluid; (ii) appearance of different environments, for example, dielectric, metallic and plasma backgrounds, or adding one more CNT; (iii) CNT placing in the waveguide or optical cavity. Such generalizations are promising from the point of increasing the force, similar as known for other types of nanostructures (see, for example, [40]). The modification of CNT molecular structure opens the ways for different applications of CNT manipulation. Among them, are: (i) using the pulling (see Fig. 4) for measurement of CNT electrochemical potential after substitutional doping of CNT [86]; using the optical force for CNT mechanical sensing [87,88]; applications of the optical force for CNT sorting [89] with respective of the length or radius.

We have developed a consistent theory for optical forces in CNTs with nonlocal conductivity (spatial dispersion). Unlike most contributions which consider 1D CNT (infinite length), it is demonstrated here that a rigorous account of CNT edge effects (finite length), can dramatically affect its optomechanical properties. The present formulation is based on implying a Fredholm integral equation of the second kind for the axial component of the surface current. The optical force is simulated by the numerical solution of an integral equation as well as by an approximate analytical solution. An excellent agreement is found between these two approaches. We demonstrate for the first time that there exists a negative optical force in a CNT (optical pulling). The pulling effect is stipulated by the non-locality of the conductivity, and it disappears in the local limit. The occurrence of optical pulling is typical for CNTs of rather small length (100nm - 200nm). For larger values of length, the influence of spatial dispersion decreases and ceases to exhibit itself at large lengths. The force becomes positive (pushing) and to linearly depend on the length. Yet another possible way to obtain an optical pulling force in a CNT, is by increasing the forward scattering by means of a structured light, such as a Gaussian beam, Bessel beam, as well as other tractor beams. It is expected that including spatial dispersion considerations of constitutive parameters, may be able to produce optical forces pulling also in other media. Our Letter paves a way for future studies of optical forces in nanostructures.

*t.berghaus@gmail.com;
**gregory_slepyan@yahoo.com;

## Supplemental Material for:

## "Optical Pulling Force in Carbon Nanotubes: Manifestation of Nonlocal Conductivity"

Tomer Berghaus, Touvia Miloh, and Gregory Ya. Slepyan

In this Supplemental Material we present some ancillary details of calculations used in this Letter. It consists of three parts; The proposed model of nonlocal CNT conductivity is presented in Part I. The derivation of the integral equation (9) for the surface current density is given in Part

II. The two different forms of the kernel of the integral equation used in calculations, are presented in Part III.

## PART I. THE USED MODEL OF NONLOCAL CONDUCTIVITY OF CNT

In this section, we consider the self-consistent mechanism of CNT conductivity based on the model of Dirac pseudospins [61]. This theory covers the wide frequency range (from THz until visible light inclusively) and takes into account the interband transitions. The appropriate Hamiltonian for this case can be written as $\hat{H} = \hat{H}_K + \hat{V}(t)$, where $\hat{H}_K = -\hbar v_F \hat{\sigma}_z \frac{\partial}{\partial z}$ represents the Hamiltonian of conductive electrons without the EM-field, ($\hbar$ is the reduced Planck's constant, $v_F$ denotes the Fermi velocity and $\hat{\sigma}_z = (0,-i;i,0)$ is the longitudinal Pauli matrix). In addition, we define $\hat{V}(t) = -eE_z\hat{z}$ ($e$ denotes the electron charge, $E_z$ is the longitudinal component of electric field and $\hat{z}$ is the operator of longitudinal coordinate). We will solve the Liouville equation for the case of a $2\times2$ pseudospin matrix $i\hbar\frac{\partial\hat{\rho}}{\partial t} = \left[\hat{H}_K,\hat{\rho}\right] + \left[\hat{V}(t),\hat{\rho}\right]$, where the second term is considered small and will be accounted for by using a perturbation technique. The formulation is based on assuming a monochromatic EM-field, which means letting $\hat{V}(t) = \mathrm{Re}\left(\hat{V}e^{-i\omega t}\right)$ and $\hat{\rho}(t) = \hat{\rho}_0 + \mathrm{Re}\left(\delta\hat{\rho}\cdot e^{-i\omega t}\right)$, where $\delta\hat{\rho}$ is taken as a small correction (perturbation) to the ambient EM-field. In addition, $\hat{\rho}_0 = F(\varepsilon) = \left[1+\exp(\varepsilon-\mu)/k_B T\right]^{-1}$ denotes the equilibrium density distribution (Fermi distribution), $\varepsilon$ is the electron's energy, $\mu$ is the electrochemical potential, $k_B$ represents the Boltzmann constant, and $T$ is the temperature.

The electromagnetic fields and the induced currents can be expressed as a superposition of traveling waves in the form of $\exp\left[i(qz-\omega t)\right]$, which propagate in opposite directions (different signs of wavenumbers $q$). For brevity, we consider here only the case of axisymmetric conductivity, which corresponds to the axial current produced by the axial component of the field. All other components of the $2\times2$ conductivity tensor, are assumed to be zero. Far from the CNT terminations (ends), their interaction can be accurately modeled by means of using the corresponding infinite-length CNT approximation. Reflections from the extreme ends of the CNT and accounting for nonlocal effects, are also incorporated by applying the additional boundary conditions [59]. Following the Kubo technique, the perturbation in the elements of the density matrix, expressed in the momentum space, can be represented as

$$(\delta\hat{\rho})_{mn,p,p+q} = \frac{F\left(\varepsilon_{m,p+q}\right) - F\left(\varepsilon_{np}\right)}{\left[\varepsilon_{np} - \varepsilon_{m,p+q} + \hbar\omega + i0\right]}\hat{V}_{mn,p,p+q} \tag{S1}$$

where $\varepsilon_{m,p}$ is the energy of the electron with momentum $p$ corresponding to a zone with index $m$. The conductivity in the momentum space reads

$$\sigma_{zz}(\omega,q) = \frac{i}{\omega}\left[\Pi(\omega,q) - \Pi(0,q)\right] \tag{S2}$$

where

$$\Pi(\omega,q)=\sum_{mn}\sum_{p}\frac{F(\varepsilon_{m,p+q})-F(\varepsilon_{np})}{\left[\varepsilon_{np}-\varepsilon_{m,p+q}-\hbar(\omega+i0)\right]}\left|\langle\hat{j}_z\rangle_{mn,p,p+q}\right|^2 \tag{S3}$$

is defined as the polarizability and $\langle\hat{j}_z\rangle_{mn,p,p+q}$ represent the matrix elements of the operators of the current density. The sum over *p in* (S3) is a superposition over the states with different momenta and is a conventional form of integration over the BZ zone. It is also convenient to split the total nonlocal conductivity into inter and intra components, namely, $\sigma_{zz}(\omega,q)=\sigma_{zz}^{\text{inter}}(\omega,q)+\sigma_{zz}^{\text{intra}}(\omega,q)$. The terms with $m\neq n$ in (S3) correspond to interband transitions while the other terms correspond to intraband transitions (also named the Drude conductivity).

To facilitate the incorporation of the nonlocal effects, we employ a Taylor series expansion of (S3) with respect to the momentum $q$, by retaining the second-order approximation (relation (2) in the paper). We will adopt the approximation $\langle\hat{j}_z\rangle_{mn,p,p+q}\approx\langle\hat{j}_z\rangle_{mn,pp}$ along with the following dispersion relation for the Dirac fermions $\varepsilon_{mp}\approx\pm v_F p$. Since the current in the CNT is supported by contributions from all available states, (S3) can also be reformulated as:

$$\Pi(\omega,q)=\sum_{mn}\sum_{p}F(\varepsilon_{np})\left|\langle\hat{j}_z\rangle_{mn,pp}\right|^2\left\{\frac{1}{\left[\varepsilon_{np}-\varepsilon_{m,p+q}-\hbar(\omega+i0)\right]}+\frac{1}{\left[\varepsilon_{np}-\varepsilon_{m,p+q}+\hbar(\omega+i0)\right]}\right\} \tag{S4}$$

Returning to the position space, we perform an inversion $q\to\partial/\partial z$ , which suggests that the total current can be expressed as the sum of local and nonlocal components, i.e., $j_z=\sigma_{zz}(\omega,0)E_z+j_{z,\text{Nonloc}}$
where

$$j_{z,\text{Nonloc}}=-\xi(\omega)\frac{\partial^2E_z}{\partial z^2} \tag{S5}$$

with the factor of nonlocality defined as

$$\xi(\omega)=\frac{1}{2}\frac{\partial^2\sigma_{zz}(\omega,q)}{\partial q^2}\bigg|_{q=0} \tag{S6}$$

and for reasons of brevity, we define $\sigma_{zz}(\omega,0)=\sigma_{zz}(\omega)$. The nonlocal component of the current may also be separated in a similar way into interband and intraband terms, thus the factor of nonlocality becomes $\xi(\omega)=\xi^{\text{inter}}(\omega)+\xi^{\text{intra}}(\omega)$. Combining (S2), (S4) and (S5) finally renders

$$\xi(\omega)=4i\hbar(\hbar v_F)^2\sum_{mn}\sum_{p}\frac{\left(F(\varepsilon_{np})-F(\varepsilon_{mp})\right)\left|\langle\hat{j}_z\rangle_{mn,pp}\right|^2}{\left[\varepsilon_{np}-\varepsilon_{mp}+\hbar(\omega+i0)\right]^3(\varepsilon_{np}-\varepsilon_{mp})}. \tag{S7}$$

Nonlocality is evident in (S5) through the presence of the second-order spatial derivative, indicating that the current density at a given point depends not only on the value of the electric field at that point, but also on its spatial vicinity. Note that the term corresponding to the first-order derivative in the Taylor series vanishes due to axial symmetry of achiral CNTs, which ensures invariance under rotation around the tube axis. Taking into account that $\partial^2\left(x^{-1}\right)/\partial x^2 = 2x^{-3}$ , we obtain

$$\frac{2\hbar^2}{\left[\varepsilon_{np}-\varepsilon_{m,p+q}\pm\hbar\omega\right]^3}=\frac{\partial^2}{\partial\omega^2}\left(\frac{1}{\varepsilon_{np}-\varepsilon_{m,p+q}\pm\hbar\omega}\right) \tag{S8}$$

and by virtue of (S8), we can rewrite (S7) in the form:

$$\xi(\omega)=-2i\hbar v_F^2\frac{\partial^2}{\partial\omega^2}\sum_{mn}\sum_{p}\frac{\left(F\left(\varepsilon_{np}\right)-F\left(\varepsilon_{mp}\right)\right)\left|\left\langle \hat{j}_z\right\rangle_{mn,pp}\right|^2}{\left[\varepsilon_{np}-\varepsilon_{mp}+\hbar\left(\omega+i0\right)\right]\left(\varepsilon_{np}-\varepsilon_{mp}\right)} \tag{S9}$$

which immediately Eq. (7) in the paper.

In order to simplify the general Kubo formalism for low-energy excitations constrained by the condition $E<1$ eV, we adapt it to the case of a pseudo-spin electron liquid, in a similar manner to the approach used for a monolayer graphene in [60]. To this end, we consider the electronic states near the K-points and employ a linear approximation for $\varepsilon_{\mathbf{p}}=\pm v_F\left|\mathbf{p}-\mathbf{p}_F\right|$ . We will also consider the case of low temperature ( $T<<\mu$ ) and a collision-free electron motion ( $\nu=0$ ). In contrast with graphene [60, 61], for a CNT we take into account the transverse electron confinement (the azimuthal component of electron momentum is discrete, while the longitudinal component is continuous).

Following [64], we refine the Kubo formalism to suit the specific model under consideration. For a zigzag CNT (*m,* 0), we have

$$\sigma_{zz}^{\text{inter}}(\omega)=\frac{ie^2\omega}{2\pi^2\hbar R_{CN}}\sum_{s=1}^{m}\int_{BZ}\frac{\left|v_{cv}\left(p_z,s\right)\right|^2}{\varepsilon\left(p_z,s\right)}\cdot\frac{F\left[-\varepsilon\left(p_z,s\right)\right]-F\left[\varepsilon\left(p_z,s\right)\right]}{\hbar^2\left(\omega+i0\right)^2-4\varepsilon^2\left(p_z,s\right)}dp_z \tag{S10}$$

where $v_{cv}\left(p_z,s\right)$ denotes a velocity matrix element of direct interband transitions [64]. For future considerations, it may be simply approximated by $v_{cv}\left(p_z,s\right)\approx\hbar v_F$ [64].

Real implemented CNTs have a finite radius, which corresponds to rather small values of *m* (*m*<60). The main contribution to the conductivity is provided by fhe Fermi points *s*=*m*/3 and *s*=2*m*/3, which correspond to the different valleys of the BZ zone. Next, we transform the integration over $p_z$ to that over energy and carry it out over the range $0<\varepsilon<\infty$ , which results in Eqs. (4), (5). For the case where $\mu>>k_BT$ (named zero-temperature approximation), the final result reads

$$\sigma_{zz}^{\text{inter}}(\omega)=\frac{e^2v_F}{2\pi\hbar\omega}\left(H\left(\hbar\omega-2\mu\right)-\frac{i}{\pi}\ln\left|\frac{4\mu^2}{4\mu^2-\left(\hbar\omega\right)^2}\right|\right) \tag{S11}$$

where *H*(*x*) is the Heaviside step function. The conductivity (S11) goes to infinity in the limit $\mu\to0$ , in accordance with the zero-temperature approximation. The nonlocality parameter $\xi$ , can be written as

$$\xi^{\text{inter}}(\omega) = i\frac{16\sigma_0 \mu v_F^2}{\pi\hbar\omega^3}\left\{1 - \frac{(\hbar\omega)^4}{\left[(\hbar\omega)^2 - 4\mu^2\right]^2}\right\} \tag{S12}$$

where $\sigma_0$ denotes the quantum of conductivity. The singularity in (S11), (S12) at the step point $\hbar\omega - 2\mu$, is removed by accounting for temperature [59].

## PART II: DERIVATION OF THE INTEGRAL EQUATION (9).

The scattering field by the CNT, can be described by the longitudinal component of the electric Hertz vector $\Pi$, as

$$E_z^{sc} = \frac{\partial^2 \Pi}{\partial z^2} + k^2 \Pi \ . \tag{S13}$$

This field is governed by the three-dimensional wave equation derived from Maxwell's equations, which can be expressed in terms of the electric Hertz vector as

$$\left(\nabla^2 + k^2\right)\Pi = -\frac{i}{\omega\varepsilon_0}\delta\left(r - R_{CN}\right) j_z \tag{S14}$$

where $\delta(x)$ is the Dirac function, and subject to effective boundary conditions prevailing on the CNT surface. The discontinuity in the tangential component of the magnetic field is proportional to the surface current density, while the tangential component of the electric field remains continuous across the surface. Denoting the interior $(R_{CN} - 0)$ and the exterior $(R_{CN} + 0)$ components of the Hertz vector by $\Pi^{(\pm)} = \Pi\left(r = R_{CN} \pm 0\right)$, the effective boundary conditions for the eigenmodes are given by

$$\Pi^{(+)} - \Pi^{(-)} = 0, \quad -\infty < z < \infty \tag{S15}$$

$$\left(1 + \frac{1}{\tilde{\alpha}^2(\omega)}\frac{\partial^2}{\partial z^2}\right)\frac{\partial}{\partial r}\left(\Pi^{(+)} - \Pi^{(-)}\right) = \frac{\sigma_{zz}(\omega)}{i\omega\varepsilon_0}\left(\frac{\partial^2}{\partial z^2} + k^2\right)\Pi + \frac{\sigma_{zz}(\omega)}{i\omega\varepsilon_0}E_z^{inc}, \quad -L < z < L \tag{S16}$$

$$\frac{\partial}{\partial r}\left(\Pi^{(+)} - \Pi^{(-)}\right) = 0, \quad |z| > L \tag{S17}$$

where the subscript $\pm$ in the right-hand part of (S16) is omitted because of continuity.

The solution of the homogeneous Helmholtz solution for the Hertz potential (S14), can be expressed as

$$j_z(z) = -\sigma_{zz}(\omega)\tilde{\alpha}^2(\omega)\int_{-L}^{L} g(z,z')E_z(z')dz' \tag{S18}$$

where $g(z,z')$ represents the corresponding Green function of Sturm-Liouville equation

$$\frac{\partial^2 g(z,z')}{\partial z^2} + \tilde{\alpha}^2(\omega)g(z,z') = \delta(z - z') \tag{S19}$$

with a Dirichlet-type boundary conditions $g(\pm L, z') = 0$. These boundary conditions guarantee satisfying the so-called 'additional' boundary conditions [59] and are necessary for the consistent (correct) description of nonlocality [59].

The Green function may be presented in the following form,

$$g(z,z') = -\frac{1}{\tilde{\alpha}(\omega)\sin(2\tilde{\alpha}(\omega)L)}[H(z'-z)\sin(\tilde{\alpha}(\omega)(z+L))\sin(\tilde{\alpha}(\omega)(z'-L)) + H(z-z')\sin(\tilde{\alpha}(\omega)(z'+L))\sin(\tilde{\alpha}(\omega)(z-L))] \quad \text{(S20)}$$

where $H(z)$ is the Heaviside function. Another form, which is based on using eigenmodes, reads

$$g(z,z') = \sum_{n=1}^{\infty} \frac{u_n(z)u_n(z')}{\left(\frac{n\pi}{2L}\right)^2 - \tilde{\alpha}^2(\omega)} \quad \text{(S21)}$$

where $u_n(z) = \sqrt{\frac{2}{L}}\sin\left(\frac{n\pi}{2L}(z+L)\right)$. In the local limit $|\tilde{\alpha}(\omega)| \to \infty$, Eq. (S21) renders $g(z,z') = -\tilde{\alpha}^{-2}(\omega)\delta(z-z')$.

Let us next express the Hertz potential as

$$\Pi(r,z) = \frac{i}{\omega\varepsilon_0}\int_{-L}^{L} j_z(s)G(r,z-s)ds \quad \text{(S22)}$$

where $j_z(z)$ represents the unknown current density and the kernel is given by

$$G(r,s-z) = R_{CN}\int_0^{2\pi} \frac{e^{ik\sqrt{r^2+R_{CN}^2-2rR_{CN}\sin(\phi)+(s-z)^2}}}{\sqrt{r^2+R_{CN}^2-2rR_{CN}\sin(\phi)+(s-z)^2}}d\phi \ . \quad \text{(S23)}$$

Substituting Eq. (S18) into Eq. (S22), yields the exact solution to Maxwell's equations throughout the entire space, by satisfying both the radiation condition and the boundary conditions expressed in Eq. (S16). The surface current density appearing in Eq. (S22) must then be determined by enforcing the boundary condition (S17). After substitution of Eq. (S22) into Eq. (S17), we arrive at the governing equation for the current density

$$j_z(z) + \frac{i\sigma_{zz}(\omega)\tilde{\alpha}^2(\omega)}{\omega\varepsilon_0}\int_{-L}^{L} g(z,z')\left(\frac{\partial^2}{\partial z'^2}+k^2\right)\left\{\int_{-L}^{L} j_z(s)G(s-z')ds\right\}dz' = B(z) \quad \text{(S24)}$$

where $G(s-z') = G(R_{CN},s-z')$ is given by Eq. (S23) and

$$B(z) = \sigma_{zz}(\omega)\tilde{\alpha}^2(\omega)\int_{-L}^{L} g(z,z')E_z^{inc}(z')dz' \ . \quad \text{(S25)}$$

Introducing the kernel

$$K(z,s) = \left(\frac{\partial^2}{\partial s^2}+k^2\right)\int_{-L}^{L} g(z,z')G(s-z')dz' \quad \text{(S26)}$$

and accounting for (S25), we finally obtain the integral equation (9).

**PART III: DIFFERENT FORMS OF THE KERNEL OF THE INTEGRAL EQUATION (9).**

**Form based on the singular integrals**

The idea of a singular integral transformation was developed in [65], in the context of wire dipole antennas. However, the integral forms for wires and CNTs are not identical because of the boundary conditions. Therefore, we make the calculations from the start and present them in detail in this subsection. Let us consider the configuration of the CNT depicted at Fig.1. The symbol $\Sigma$ denotes the surface of the CNT ( $d\Sigma = 2\pi R_{CN} dz d\phi$ is the differential of this surface). The symbol $\Sigma_\delta$ stands for an exclusion element, which simplifies the transformation of the singular integrals, by excluding the singularities. The symbols $C, C_\delta$ represent the two-side closed rims of the CNT and the excluded element, respectively. Next, we consider the integral

$$v(z') = \lim_{\delta \to 0} \int_{\Sigma - \Sigma_\delta} j_z(s) \left( \frac{e^{ikR}}{R} \right) d\Sigma \tag{S27}$$

where $R = \sqrt{4R_{CN}^2 \sin^2\left(\frac{\phi}{2}\right) + (s - z')^2}$ .

Thus,

$$\begin{aligned}
\frac{\partial v(z')}{\partial z'} &= \lim_{\delta \to 0} \int_{\Sigma - \Sigma_\delta} j_z(s) \frac{\partial}{\partial z'} \left( \frac{e^{ikR}}{R} \right) d\Sigma = \\
&-\lim_{\delta \to 0} \int_{\Sigma - \Sigma_\delta} j_z(s) \frac{\partial}{\partial s} \left( \frac{e^{ikR}}{R} \right) d\Sigma = \\
&\lim_{\delta \to 0} \left[ \int_{\Sigma - \Sigma_\delta} \frac{\partial j_z(s)}{\partial s} \left( \frac{e^{ikR}}{R} \right) d\Sigma - \int_{\Sigma - \Sigma_\delta} \frac{\partial}{\partial s} \left( j_z(s) \frac{e^{ikR}}{R} \right) d\Sigma \right] = \\
&\lim_{\delta \to 0} \left[ \int_{\Sigma - \Sigma_\delta} \frac{\partial j_z(s)}{\partial s} \left( \frac{e^{ikR}}{R} \right) d\Sigma - \int_C j_z(s) \frac{e^{ikR}}{R} dC \right]
\end{aligned} \tag{S28}$$

where the integral over $C_\delta$ in the last term vanishes. Performing the second derivative gives,

$$\begin{aligned}
\frac{\partial^2 v(z')}{\partial z'^2} &= \lim_{\delta \to 0} \left[ \int_{\Sigma - \Sigma_\delta} \frac{\partial \left( j_z(s) - j_z(z') \right)}{\partial z'} \frac{\partial}{\partial z'} \left( \frac{e^{ikR}}{R} \right) d\Sigma - \int_C j_z(s) \frac{\partial}{\partial z'} \left( \frac{e^{ikR}}{R} \right) dC \right] = \\
&\lim_{\delta \to 0} \left[ \int_{\Sigma - \Sigma_\delta} \frac{\partial \left( j_z(s) - j_z(z') \right)}{\partial z'} \frac{\partial^2}{\partial z'^2} \left( \frac{e^{ikR}}{R} \right) d\Sigma + \int_C \left( j_z(s) - j_z(z') \right) \frac{\partial}{\partial z'} \left( \frac{e^{ikR}}{R} \right) dC \right]
\end{aligned} \tag{S29}$$

Finally by using

$$\begin{aligned}
&\lim_{\delta \to 0} \int_{\Sigma - \Sigma_\delta} \left( j_z(s) - j_z(z') \right) \frac{\partial^2}{\partial z'^2} \left( \frac{e^{ikR}}{R} \right) d\Sigma = \\
&V.P. \int_{-L}^{L} \frac{\partial^2}{\partial s^2} \left( \int_0^{2\pi} \frac{e^{ikR}}{R} d\phi \right) \left( j_z(s) - j_z(z') \right) dz'
\end{aligned} \tag{S30}$$

and

$$\int_C \left(j_z(s)-j_z(z')\right)\frac{\partial}{\partial z'}\left(\frac{e^{ikR}}{R}\right)dC = \\ j_z(z')\left(\int_0^{2\pi}\frac{e^{ikR}}{R}d\phi\bigg|_{z'=-L} - \int_0^{2\pi}\frac{e^{ikR}}{R}d\phi\bigg|_{z'=L}\right) \quad \text{(S31)}$$

we obtain

$$\frac{\partial^2 v(z')}{\partial z'^2} = V.P.\int_{-L}^{L}\frac{\partial^2 G(z'-s)}{\partial s^2}\left(j_z(s)-j_z(z')\right)ds + \\ j_z(z')\frac{\partial}{\partial z'}\left(G(z'-L)-G(z'+L)\right) \quad \text{(S32)}$$

Multiplying (S32) by $g(z,z')$ and integrating over the CNT surface, one gets

$$\int_{-L}^{L}\frac{\partial^2 v(z')}{\partial z'^2}g(z,z')dz' = V.P.\int_{-L}^{L} j_z(s)\int_{-L}^{L}\frac{\partial^2 G(z'-s)}{\partial s^2}g(z,z')dz'ds - \\ -V.P.\int_{-L}^{L} j_z(z')\int_{-L}^{L}\frac{\partial^2 G(z'-s)}{\partial s^2}g(z,z')dsdz' + \int_{-L}^{L} j_z(z')\frac{\partial}{\partial z'}\left(G(z'-L)-G(z'+L)\right)dz' \quad \text{(S33)}$$

Note that in the second term on the right-hand side of (S33), we have exchanged $\partial^2/\partial s^2 \to \partial^2/\partial z'^2$ , $s \to z', z' \to s$ , and $z' \to s$ in the third term. As a result, we can express the kernel in a convenient form as

$$K(z,s) = \left(\frac{\partial^2}{\partial s^2}+k^2\right)\int_{-L}^{L} g(z,z')G(s-z')dz' = F_1(z,s)-F_2(z,s)+F_3(z,s) \quad \text{(S34)}$$

$$F_1(z,s) = k^2\text{V.P.}\int_{-L}^{L} g(z,z')G(s-z')dz' \quad \text{(S35)}$$

$$F_2(z,s) = g(z,s)\frac{\partial}{\partial s}\left(G(s-L)-G(s+L)\right) \quad \text{(S36)}$$

$$F_3(z,s) = \text{V.P.}\int_{-L}^{L}\left(g(z,z')-g(z,s)\right)\frac{\partial^2 G(s-z')}{\partial s^2}dz' \quad \text{(S37)}$$

where the symbol V.P. denotes the Cauchy principal value of the integral. This is one of the forms of the kernel, which are used in our numerical simulations.

**Form based on the additional theorem for the Hankel function**

Expanding the Green function of the Helmholtz equation by cylindrical modes, yields

$$\frac{e^{ik\sqrt{r^2+R_{CN}^2-2rR_{CN}\sin(\phi)+(s-z)^2}}}{\sqrt{r^2+R_{CN}^2-2rR_{CN}\sin(\phi)+(s-z)^2}} = \frac{i}{2}\int_{-\infty}^{\infty} e^{ih(z-s)}H_0^{(1)}\left(\sqrt{k^2-h^2}\bar{r}\right)dh \quad \text{(S38)}$$

where $\overline{r} = \sqrt{r^2 + R_{CN}^2 - 2rR_{CN}\sin(\phi)}$. Using the additional theorem for the Hankel functions, implies

$$H_0^{(1)}(\kappa\overline{r}) = \sum_n H_{-n}^{(1)}(\kappa r) J_n(\kappa R_{CN}) e^{in\phi} \tag{S39}$$

where $r > R_{CN}$. Substituting (S38) and (S39) in Eq. (12), integrating over the angle $\phi$ ,keeping only one term with $n = 0$, gives

$$G(z-s) = G(R_{CN}, z-s) = i\pi R_{CN} \int_{-\infty}^{\infty} e^{ih(z-s)} H_0^{(1)}(\kappa R_{CN}) J_0(\kappa R_{CN}) dh \tag{S40}$$

where $\kappa = \sqrt{k^2 - h^2}$. Restricting the sign of $\mathrm{Im}\left(\sqrt{k^2 - h^2}\right)$ is imposed to assure the proper decay of the Hankel function as $|h| \to \infty$.

As a next step, we substitute (S.40) in Eqs. (S35) - (S37), which leads to

$$K(z,s) = i\pi R_{CN}\left(\frac{\partial^2}{\partial s^2} + k^2\right)\int_{-\infty}^{\infty} e^{ihs} H_0^{(1)}(\kappa R_{CN}) J_0(\kappa R_{CN})\left(\int_{-L}^{L} g(z,z') e^{-ihz'} dz'\right) dh \tag{S41}$$

where $g(z,z')$ is the Green function given in (S20). An integration over $z'$ readily yields

$$\int_{-L}^{L} g(z,z') e^{-ihz'} dz' = -\frac{F(h,z)}{(\tilde{\alpha}^2 - h^2)\sin(2\tilde{\alpha}L)} \tag{S42}$$

where $F(h,z) = e^{-ihz}\sin(2\tilde{\alpha}L) + e^{ihL}\sin X - e^{-ihL}\sin Y$ , and $X = \tilde{\alpha}(z-L), Y = \tilde{\alpha}(z+L)$. The right-hand side of (S42) is nonsingular at $h = \pm\tilde{\alpha}$, since $F(\pm\tilde{\alpha}, z) = 0$. Note that this function also satisfies the following symmetry relation $F(h,z) = F(-h,-z)$, which implies that $K(z,s) = K(-z,-s)$ .Finally, by changing the order of integration over *h* in (S41), and substituting (S42) in (S41), we obtain

$$K(z,s) = -\frac{i\pi R_{CN}}{\sin(2\tilde{\alpha}L)} \int_{-\infty}^{\infty} e^{ihs}\left(k^2 - h^2\right) \frac{H_0^{(1)}(\kappa R_{CN}) J_0(\kappa R_{CN})}{\kappa^2 - k^2} F(z,h) dh \tag{S43}$$

This is an another form of the kernel, which is equivalent to the form (S34) that leads to the same numerical results.